\documentclass[11pt]{article}
\usepackage{latexsym}
\usepackage{amssymb}
\usepackage[dvips]{graphicx}
\usepackage{epsfig}
\usepackage{rotating}
\usepackage{amsfonts}
\usepackage{amsmath}
\begin{document}

\begin{titlepage}
\begin{flushright}
CP3-09-41\\
\end{flushright}

\vspace{40pt}

\begin{center}

{\Large\bf Variations on the Planar Landau Problem:}

\vspace{5pt}

{\Large\bf Canonical Transformations, A Purely Linear Potential}

\vspace{5pt}

{\Large\bf and the Half-Plane}\\

\vspace{20pt}

Jan Govaerts$^{a,b,}$\footnote{Fellow of the Stellenbosch Institute
for Advanced Study (STIAS), 7600 Stellenbosch, South Africa.},
M. Norbert Hounkonnou$^{b}$ and Habatwa V. Mweene$^{c}$

\vspace{15pt}

$^{a}${\sl Center for Particle Physics and Phenomenology (CP3),\\
Institut de Physique Nucl\'eaire, Universit\'e catholique de Louvain (U.C.L.),\\
2, Chemin du Cyclotron, B-1348 Louvain-la-Neuve, Belgium}\\
E-mail: {\em Jan.Govaerts@uclouvain.be}\\
\vspace{10pt}
$^{b}${\sl International Chair in Mathematical Physics and Applications (ICMPA--UNESCO Chair),\\
University of Abomey--Calavi, 072 B. P. 50, Cotonou, Republic of Benin}\\
E-mail: {\em hounkonnou@yahoo.fr, norbert.hounkonnou@cipma.uac.bj}\\
\vspace{10pt}
$^{c}${\sl Physics Department, University of Zambia\\
P. O. Box 32379, Lusaka, Zambia}\\
E-mail: {\em habatwamweene@yahoo.com, hmweene@unza.zm } 

\vspace{20pt}


\vspace{10pt}

\begin{abstract}
\noindent
The ordinary Landau problem of a charged particle in a plane subjected to a perpendicular homogeneous and
static magnetic field is reconsidered from different points of view. The r\^ole of phase space
canonical transformations and their relation to a choice of gauge in the solution of the problem is
addressed. The Landau problem is then extended to different contexts, in particular the singular
situation of a purely linear potential term being added as an interaction, for which
a complete purely algebraic solution is presented. This solution is then exploited to solve this same singular
Landau problem in the half-plane, with as motivation the potential relevance of such a geometry
for quantum Hall measurements in the presence of an electric field or a gravitational quantum well.
\end{abstract}

\end{center}

\end{titlepage}

\setcounter{footnote}{0}

\section{Introduction}
\label{Intro}

The classic textbook example\cite{landau} of the quantum Landau problem has remained a constant source of fascination
and inspiration\cite{jackiw}, in fields apparently so diverse as two dimensional collective quantum fermionic systems\cite{suss,jel},
the search towards a fundamental unification of gravity with the other quantum interactions, or noncommutative deformation
quantisation of geometries\cite{scholtz,scholtz2}. The same algebraic structures are also realised in M-theory
in specific limits of some background field configurations\cite{SW}. It is in view of the latter developments as well as
the phenomenology of the integer and fractional quantum Hall effects that
in recent years the Landau problem has become once again the focus of intense interest.

Yet, there remain somewhat intriguing issues open even for the simple original Landau problem. Consider thus a charged particle
of mass $m$ moving in an Euclidean plane of coordinates $(x_1,x_2)$ and subjected to a static and homogeneous magnetic
field perpendicular to that plane, with a component $B$ along the right-handed perpendicular direction which, without
loss of generality (by choosing the plane orientation appropriately), may be taken to be positive, $B>0$ (this factor, $B$,
is also normalised so as to absorb the charge of the particle). Denoting by $(A_1(x_1,x_2),A_2(x_1,x_2))$ the components
of a vector potential from which the magnetic field derives, $\partial_1 A_2 - \partial_2 A_1=B$,
it is well known that the dynamics of the system is
specified through the variational principle from the following Lagrange function,
\begin{equation}
L=\frac{1}{2}m\left(\dot{x}^2_1+\dot{x}^2_2\right)+\dot{x}_1 A_1(x_1,x_2) + \dot{x}_2 A_2(x_1,x_2),
\label{eq:L1}
\end{equation}
with as Hamiltonian function for the canonically conjugate pairs of phase space variables $(x_i,p_i)$ ($i=1,2$),
\begin{equation}
H=\frac{1}{2m}\Big(p_1-A_1(x_1,x_2)\Big)^2+\frac{1}{2m}\Big(p_2-A_2(x_1,x_2)\Big)^2.
\end{equation}

The usual discussion\cite{landau} considers the Landau gauge for the vector potentiel,
\begin{equation}
A^{\rm Landau}_1=-Bx_2,\qquad A^{\rm Landau}_2=0,
\end{equation}
in which case one has,
\begin{equation}
H=\frac{1}{2m}p^2_2+\frac{1}{2}m\omega^2_c\Big(x_2+\frac{1}{B}p_1\Big)^2,
\label{eq:HLandau}
\end{equation}
with the cyclotron frequency $\omega_c=B/m$. For the quantised system, by introducing the Fock algebra
generators
\begin{equation}
a=\sqrt{\frac{m\omega_c}{2\hbar}}\left[\left(\hat{x}_2+\frac{1}{B}\hat{p}_1\right)+\frac{i}{m\omega_c}\hat{p}_2\right],\qquad
a^\dagger=\sqrt{\frac{m\omega_c}{2\hbar}}\left[\left(\hat{x}_2+\frac{1}{B}\hat{p}_1\right)-\frac{i}{m\omega_c}\hat{p}_2\right],
\label{eq:FockLandau}
\end{equation}
with,
\begin{equation}
\Big[\hat{x}_1,\hat{p}_1\Big]=i\hbar\mathbb{I},\qquad
\Big[a,\hat{p}_1\Big]=0,\qquad
\Big[a,\hat{x}_1\Big]=-i\sqrt{\frac{\hbar}{2B}}\mathbb{I},\qquad
\Big[a,a^\dagger\Big]=\mathbb{I},
\end{equation}
such that $\hat{H}=\hbar\omega_c\left(a^\dagger a+1/2\right)$, it is clear that the energy spectrum is spanned
by Fock states $|n,p_1\rangle$ ($n=0,1,2,\ldots$) with an energy $\hbar\omega_c(n+1/2)$ which is degenerate in $p_1$---the famous Landau
levels---, the latter real variable $p_1$ labelling the $\hat{p}_1$ eigenstates.

However what is puzzling, perhaps, about this solution is the fact that because of the free particle plane wave
component of the configuration space wave function representation of these states related to the $p_1$ eigenvalue,
none of these states is normalisable,
\begin{equation}
\langle n,p_1|m,p'_1\rangle=\delta_{nm}\,\delta(p_1 - p'_1),
\end{equation}
while this basis of states is non countable and their wave functions are localised
only in the $x_2$ direction (through the Gaussian factor and Hermite
polynomials in the $(x_2+p_1/B)$ variable) but not at all in the $x_1$ where they display complete delocalisation
(note also that the $(\hat{x}_1,\hat{p}_1)$ sector does not commute with the Fock algebra, only the conjugate momentum
operator, $\hat{p}_1$, does). And yet the classical trajectories of such a particle are
circles of which the radius is function of the energy of the solution, the angular frequency is $\omega_c$, and
the magnetic center is pinned at a static position in the plane dependent on the initial conditions.
Hence rather than the above quantum states, one should expect there ought to exist another basis of the
energy eigenstates which describes normalisable and localised wave functions.

Indeed as is well known, in the circular or symmetric gauge,
\begin{equation}
A^{\rm sym}_1=-\frac{1}{2}Bx_2,\qquad A^{\rm sym}_2=+\frac{1}{2}Bx_1,
\label{eq:symgauge}
\end{equation}
once expanded, the Hamiltonian,
\begin{equation}
H=\frac{1}{2m}\left(p_1+\frac{1}{2}Bx_2\right)^2+\frac{1}{2m}\left(p_2-\frac{1}{2}Bx_1\right)^2,
\end{equation}
coincides with that of a two dimensional spherically symmetric harmonic oscillator of angular frequency
$\omega_c/2$ to which a term proportional to its angular-momentum is added. Working in a complex parametrisation
of the plane, it is clear\footnote{This is detailed in Section~\ref{Sect2}.} that the system is then diagonalised
with a countable energy eigenspectrum of Fock states, possessing the same energy spectrum as above, but now represented
by wave functions which are all localised and normalisable (and in fact centered onto the point $(x_1,x_2)=(0,0)$).

At first sight, what distinguishes the above two gauge choices at the quantum level is a redefinition of
the wave functions of quantum states by a pure phase factor, $e^{i\chi(x_1,x_2)}$, related to the gauge transformation mapping the
two choices of vector potentials into one another,
\begin{equation}
A^{\rm sym}_i=A^{\rm Landau}_i+\partial_i\chi,\qquad
\chi(x_1,x_2)=\frac{1}{2}B x_1 x_2.
\end{equation}
The phase factor $e^{i\chi}$ being singular at the point at infinity in the plane, could be thought to be the reason
for the non normalisability and non localisability of energy eigenstates in the Landau gauge. However, being a pure phase,
such a phase redefinition alone cannot explain why out of a localised and normalised wave function in the
symmetric gauge one obtains a delocalised and non normalisable one in the Landau gauge.

In Section~\ref{Sect2} this question is addressed in detail, and resolved. Then in Section~\ref{Sect3}, using
the understanding gained from Section~\ref{Sect2}, and mostly to set notations for later use, the original
Landau problem is extended by adding an interaction potential energy which combines that of a spherically
symmetric harmonic well and a linear potential. When the harmonic well is removed an apparent puzzle arises,
the resolution of which is discussed in Section~\ref{Sect4} using a purely algebraic approach.
To the best of our knowledge, the present algebraic solution---as opposed to a wave function solution
of the Schr\"odinger equation in the Landau gauge---for the Landau problem extended with a linear potential
is not available in the literature. Finally, using the insight provided by this construction and motivated by
some physical considerations, Section~\ref{Sect5} discusses the same linearly extended Landau problem in the half-plane.
The paper ends with some Conclusions.

\section{The Ordinary Landau Problem}
\label{Sect2}

\subsection{A general choice of gauge}
\label{Sect2.1}

With the Lagrangian defined in (\ref{eq:L1}), let us consider the following general class of gauge choices
for the vector potential,
\begin{equation}
A_1(x_1,x_2)=-\frac{1}{2}B\left(x_2-\overline{x}_2\right)\,+\,\partial_1\chi(x_1,x_2),\qquad
A_2(x_1,x_2)=\frac{1}{2}B\left(x_1-\overline{x}_1\right)\,+\,\partial_2\chi(x_1,x_2).
\label{eq:generalgauge}
\end{equation}
Here $(\overline{x}_1,\overline{x}_2)$ are two constant parameters representing the position
of a particular point in the plane, about which configuration space wave functions representing
the Fock states to be identified hereafter are centered and localised. Furthermore, $\chi(x_1,x_2)$ is
an arbitrary real function representing a possible gauge redefinition of the chosen vector potential.
Note that the parameters $(\overline{x}_1,\overline{x}_2)$ could also be absorbed into that gauge
transformation function, but it is useful to keep these two constants explicit.
Clearly the previous symmetric gauge corresponds to the values $(\overline{x}_1,\overline{x}_2)=(0,0)$ and
$\chi=0$, while the Landau gauge to $(\overline{x}_1,\overline{x}_2)=(0,0)$ and $\chi=-Bx_1 x_2/2$.

Incidentally, it may easily be checked that the Euler--Lagrange equations of motion that derive from
(\ref{eq:L1}) are gauge invariant, namely independent both from $(\overline{x}_1,\overline{x}_2)$ and
$\chi(x_1,x_2)$, as it should of course.

For what concerns the classical Hamiltonian formulation of the
system, the Hamiltonian reads,
\begin{equation}
H=\frac{1}{2m}\left(p_1+\frac{1}{2}B\left(x_2-\overline{x}_2\right)-\partial_1\chi\right)^2\,+\,
\frac{1}{2m}\left(p_2-\frac{1}{2}B\left(x_1-\overline{x}_1\right)-\partial_2\chi\right)^2,
\end{equation}
where the phase space variables $(x_i,p_i)$ possess canonical Poisson brackets, $\left\{x_i,p_j\right\}=\delta_{ij}$
($i,j=1,2$). Introducing now the following new parametrisation of phase space,
\begin{equation}
u_i=x_i-\overline{x}_i,\qquad
\pi_i=p_i-\partial_i\chi(x_i),
\end{equation}
which defines yet again canonically conjugate pairs of variables,
\begin{equation}
\Big\{u_i,u_j\Big\}=0,\qquad
\Big\{u_i,\pi_j\Big\}=\delta_{ij},\qquad
\Big\{\pi_i,\pi_j\Big\}=0,
\end{equation}
one has,
\begin{eqnarray}
H &=& \frac{1}{2m}\left(\pi_1+\frac{1}{2}Bu_2\right)^2\,+\,\frac{1}{2m}\left(\pi_2-\frac{1}{2}Bu_1\right)^2 \nonumber \\
&=& \frac{1}{2m}\left(\pi^2_1+\pi^2_2\right)+\frac{1}{2}m\frac{\omega^2_c}{4}\left(u^2_1+u^2_2\right)-
\frac{1}{2}\omega_c\left(u_1\pi_2 - u_2\pi_1\right).
\end{eqnarray}
The system has thereby been brought into the form it has in the symmetric gauge centered at $(x_1,x_2)=(0,0)$,
independently of the original choice of gauge. Note well that this includes the Landau gauge, however now with
a choice of phase space canonical coordinates which differs from that which led to (\ref{eq:HLandau}). This
point is addressed more specifically hereafter.

The resolution of the quantised system is now straightforward. Given the quantum commutation relations,
\begin{equation}
\Big[\hat{u}_i,\hat{\pi}_j\Big]=i\hbar\delta_{ij}\,\mathbb{I},\qquad
\hat{u}^\dagger_i=\hat{u}_i,\qquad
\hat{\pi}^\dagger_i=\hat{\pi}_i,
\end{equation}
one first introduces the cartesian Fock algebra generators,
\begin{equation}
a_i=\frac{1}{2}\sqrt{\frac{B}{\hbar}}\left(\hat{u}_i+\frac{2i}{B}\hat{\pi}_i\right),\qquad
a^\dagger_i=\frac{1}{2}\sqrt{\frac{B}{\hbar}}\left(\hat{u}_i-\frac{2i}{B}\hat{\pi}_i\right),
\end{equation}
such that
\begin{equation}
\Big[ a_i,a^\dagger_j\Big]=\delta_{ij}\,\mathbb{I},
\end{equation}
while,
\begin{equation}
\hat{H}=\frac{1}{2}\hbar\omega_c\left(a^\dagger_1 a_1\,+\, a^\dagger_2 a_2\,+\,1\right)\,+\,
\frac{1}{2}i\hbar\omega_c\left(a^\dagger_1 a_2 - a^\dagger_2 a_1\right).
\end{equation}
Next one introduces the chiral Fock algebra generators,
\begin{equation}
a_\pm=\frac{1}{\sqrt{2}}\left(a_1 \mp i a_2\right),\qquad
a^\dagger_\pm=\frac{1}{\sqrt{2}}\left(a^\dagger_1 \pm i a^\dagger_2\right),
\end{equation}
such that,
\begin{equation}
\Big[ a_\pm, a^\dagger_\pm\Big]=\mathbb{I},\qquad
\Big[ a_\pm, a^\dagger_\mp\Big]=0.
\end{equation}
A direct substitution\footnote{The inverse relations expressing $\hat{u}_i$ and $\hat{\pi}_i$ in terms of
$(a_\pm,a^\dagger_\pm)$ are easily worked out.} then finds,
\begin{equation}
\hat{H}=\hbar\omega_c\left(a^\dagger_- a_-+\frac{1}{2}\right).
\end{equation}
Consequently, given the orthonormalised Fock state basis $|n_-,n_+\rangle$ ($n_\pm=0,1,2,\ldots$) defined by
\begin{equation}
|n_-,n_+\rangle=\frac{1}{\sqrt{n_-!\,n_+!}}\left(a^\dagger_-\right)^{n_-}
\left(a^\dagger_+\right)^{n_+}\,|0\rangle,\qquad a_\pm|0\rangle=0,\qquad \langle 0|0\rangle=1,
\end{equation}
these states diagonalise the energy eigenspectrum of the system,
\begin{equation}
\hat{H}|n_-,n_+\rangle = E(n_-)\,|n_-,n_+\rangle,\qquad
E(n_-)=\hbar\omega_c\left(n_- + \frac{1}{2}\right).
\end{equation}

Hence indeed the same energy spectrum as in the Landau gauge is obtained, however now with a countable basis of
eigenstates which are all normalisable and localised in the plane. More specifically, it may be shown\cite{Lee} that
in the configuration space representation the wave functions of these chiral Fock states are given as,
\begin{equation}
\langle x_1,x_2|n_-,n_+\rangle=\frac{(-1)^m}{\sqrt{2\pi\hbar}}\,
\sqrt{\frac{m!}{(m+|\ell|)!}}\,u^{|\ell|/2}\,e^{i\ell\theta}\,e^{-\frac{1}{2}u}\,L^{|\ell|}_m(u),
\label{eq:Laguerre}
\end{equation}
where $\ell=n_+ - n_-$, $m={\rm min}(n_-,n_+)=n_-+(\ell-|\ell|)/2$ and $L^{|\ell|}_m(u)$ are the generalised
Laguerre polynomials, while,
\begin{equation}
u=\frac{m\omega_c}{2\hbar}\left[\left(x_1-\overline{x}_1\right)^2+\left(x_2-\overline{x}_2\right)^2\right],\qquad
e^{i\theta}=\frac{\left(x_1-\overline{x}_1\right)+i\left(x_2-\overline{x}_2\right)}
{\sqrt{\left(x_1-\overline{x}_1\right)^2+\left(x_2-\overline{x}_2\right)^2}}.
\end{equation}
Clearly all these states are thus indeed localised and centered at the point $(\overline{x}_1,\overline{x}_2)$,
and normalisable, independently of the chosen gauge for the vector potential, including the Landau gauge.
This result is achieved by having identified the appropriate canonical phase space transformation which
undoes any gauge transformation away from the symmetric gauge, while at the same time moving the set of
localised Fock states to be centered at any given point in the plane.

\subsection{The solution in the Landau gauge}
\label{Sect2.2}

In terms of the general parametrisation for a gauge choice in (\ref{eq:generalgauge}), the Landau
gauge as defined in the Introduction corresponds to the function
\begin{equation}
\chi(x_1,x_2)=-\frac{1}{2}B\left(x_1-\overline{x}_1\right)\left(x_2+\overline{x}_2\right).
\end{equation}
Consequently, one then finds,
\begin{eqnarray}
\pi_1 &=& p_1+\frac{1}{2}B\left(x_2+\overline{x}_2\right),\qquad
\pi_1+\frac{1}{2}Bu_2=\pi_1+\frac{1}{2}B\left(x_2-\overline{x}_2\right)=p_1+B x_2, \nonumber \\
\pi_2 &=& p_2+\frac{1}{2}B\left(x_1-\overline{x}_1\right),\qquad
\pi_2-\frac{1}{2}Bu_1=\pi_2-\frac{1}{2}B\left(x_1-\overline{x}_1\right)=p_2,
\end{eqnarray}
so that indeed,
\begin{equation}
H=\frac{1}{2m}\left(p_1+Bx_2\right)^2+\frac{1}{2m}p^2_2.
\end{equation}

Given these relations in the Landau gauge, it is now possible to express the operators
$\hat{x}_i$ and $\hat{p}_i$ in terms of the cartesian and chiral Fock operators introduced above. One then finds,
\begin{equation}
\hat{x}_1=\overline{x}_1+\sqrt{\frac{\hbar}{B}}\left(a_1+a^\dagger_1\right),\qquad
\hat{x}_2=\overline{x}_2+\sqrt{\frac{\hbar}{B}}\left(a_2+a^\dagger_2\right),
\end{equation}
\begin{eqnarray}
\hat{p}_1 &=& -\frac{1}{2}i\sqrt{\hbar B}\left(a_1-a^\dagger_1\right)-\frac{1}{2}\sqrt{\hbar B}\left(a_2+a^\dagger_2\right)
-B\overline{x}_2, \nonumber \\
\hat{p}_2 &=& -\frac{1}{2}i\sqrt{\hbar B}\left(a_2 - a^\dagger_2\right)-\frac{1}{2}\sqrt{\hbar B}\left(a_1 + a^\dagger_1\right).
\end{eqnarray}
We then have,
\begin{equation}
\hat{p}_2=-\sqrt{\frac{\hbar B}{2}}\left(a_- + a^\dagger_-\right),\qquad
\hat{x}_2+\frac{1}{B}\hat{p}_1=-i\sqrt{\frac{\hbar}{2B}}\left(a_- - a^\dagger_-\right),
\end{equation}
so that the $(a,a^\dagger)$ Fock generators defined in (\ref{eq:FockLandau}) in the Landau gauge correspond to,
\begin{equation}
a=-i a_-,\qquad a^\dagger=i a^\dagger_-.
\end{equation}
Hence we have indeed that
\begin{equation}
\hat{H}=\hbar\omega_c\left(a^\dagger_- a_- +\frac{1}{2}\right)=\hbar\omega_c\left(a^\dagger a +\frac{1}{2}\right),
\end{equation}
explaining why the same values for the energy spectrum are obtained in both constructions for the quantum solution.
However, in the discussion as recalled in the Introduction the degeneracy of Landau levels is accounted for in terms
of the eigenstates of $\hat{p}_1$, namely,
\begin{equation}
\hat{p}_1=-i\sqrt{\frac{\hbar B}{2}}\left(a_+ - a^\dagger_+\right)\,-\,B\overline{x}_2,
\end{equation}
rather than the Fock states $|n_+\rangle$ of the Fock algebra $(a_+,a^\dagger_+)$ as obtained in the
previous general solution irrespective of the choice of gauge. Therefore when
expressed in terms of these Fock states, the solution to the eigenvalue equation,
\begin{equation}
\hat{p}_1|p_1\rangle=p_1 |p_1\rangle,\qquad p_1\in\mathbb{R},
\end{equation}
involves an infinite linear combination of all Fock states $|n_+\rangle$ which is not normalisable.

In other words, the reason why the usual solution to the Landau problem in the Landau gauge leads
to states which, within each of the Landau levels, are not normalisable nor localised, is not at all
related to that particular choice of gauge. Rather, it is because that choice of gauge naturally leads
one to use such a canonical parametrisation of phase space which upon its canonical quantisation produces
a basis of energy eigenstates which, in each Landau level, are not normalisable nor localised.
However this singular character in the choice of basis within each Landau level is avoided by
an appropriate canonical transformation which upon its canonical quantisation produces a basis of
energy eigenstates which, as Fock states, are all normalisable and localised irrespective of the choice of gauge.
It is thus coincidental
that precisely in the Landau gauge, the generic canonical phase space parametrisation valid for
any choice of gauge is just not manifest enough, so that one is lead rather onto a path towards
another construction of a solution for energy eigenstates which are no longer normalisable nor localised.

This analysis thus also shows that it is preferable to consider in all cases a parametrisation of the
general choice of gauge as in (\ref{eq:generalgauge}), which in effect is a gauge transformed form of the
symmetric gauge. One is then assured that if energy eigenstates are not normalisable or localised, there
is actually a physical justification or meaning to such a singular character, rather than being due
to some inappropriate choice of canonical parametrisation of phase space.

\section{The Landau Problem with a Quadratic and Linear Potential}
\label{Sect3}

Given the above understanding of the preferred choice of gauge, let us now consider an extension
of the Landau problem which includes an interaction potential energy, $V(x_1,x_2)$, still leading
to linear equations of motion, whether at the classical level or the quantum one in the Heisenberg
picture. In order to remain consistent with the rotational invariance of the original problem,
this potential consists of a spherically symmetric harmonic well of angular frequency $\omega_0>0$
centered at the origin $(x_1,x_2)=(0,0)$, to which a linear term is added, lying---by an appropriate choice
of planar coordinates $(x_1,x_2)$---in the $x_2$ direction,
\begin{equation}
V(x_1,x_2)=\frac{1}{2}m\omega^2_0\left(x^2_1+x^2_2\right)+\gamma x_2.
\end{equation}
Here $\gamma$ is a real constant parameter, setting the strength of a constant pull onto the particle
in the $(-x_2)$ direction (for positive $\gamma$). This linear potential may correspond, for instance, to a constant electric
field lying inside the plane and along the $x_2$ direction. Another possibility is a gravitational
potential term if the plane is tilted with respect to the horizontal direction by some angle $\alpha$,
in which case one has $\gamma=mg\cos\alpha$ if $x_2$ increases up-wards, $g>0$ being the gravitational acceleration.
These two examples thus indicate to which types of physical configurations such a linear potential could correspond.

Choosing for the vector potential the symmetric gauge as in (\ref{eq:symgauge}) does not lead to
a straightforward resolution of either the Hamiltonian or the quantum dynamics. Indeed, since that choice
is centered onto the point $(\overline{x}_1,\overline{x}_2)=(0,0)$, it clashes with the fact that
because of the linear term in the potential energy, the total potential energy---still spherically
symmetric---is centered onto a minimal position given by,
\begin{equation}
\overline{x}_1=0,\qquad
\overline{x}_2=-\frac{\gamma}{m\omega^2_0},
\label{eq:center2}
\end{equation}
since,
\begin{equation}
V(x_1,x_2)=\frac{1}{2}m\omega^2_0\left(x^2_1+\left(x_2+\frac{\gamma}{m\omega^2_0}\right)^2\right)\,-\,
\frac{\gamma^2}{2m\omega^2_0}.
\end{equation}
Obviously classical trajectories will then be centered onto that static average position in the plane.
Consequently, it is preferable to adapt the choice of symmetric gauge in the following way,
\begin{equation}
A_1(x_1,x_2)=-\frac{1}{2}B\left(x_2+\frac{\gamma}{m\omega^2_0}\right),\qquad
A_2(x_1,x_2)=+\frac{1}{2}Bx_1.
\end{equation}

The Hamiltonian is then of the form,
\begin{eqnarray}
H &=& \frac{1}{2m}\left(p_1+\frac{1}{2}B\left(x_2+\frac{\gamma}{m\omega^2_0}\right)\right)^2
+\frac{1}{2m}\left(p_2-\frac{1}{2}Bx_1\right)^2+\frac{1}{2}m\omega^2_0\left(x^2_1+x^2_2\right)+\gamma x_2  \nonumber \\
&=& \ \ \frac{1}{2m}\left(p^2_1+p^2_2\right)+\frac{1}{2}m\omega^2\left(x^2_1+\left(x_2+\frac{\gamma}{m\omega^2_0}\right)^2\right) \nonumber \\
&& -\frac{1}{2}\omega_c\left(x_1 p_2 - \left(x_2+\frac{\gamma}{m\omega^2_0}\right)p_1\right)-\frac{\gamma^2}{2m\omega^2_0},
\end{eqnarray}
where $\omega=\sqrt{\omega^2_0+\omega^2_c/4}$.

The diagonalisation of this quantum Hamiltonian is now straightforward enough\footnote{Had one not chosen
the symmetric gauge centered onto the point in (\ref{eq:center2}), there would have remained terms linear in $\hat{p}_1$
in $\hat{H}$ spoiling the simplicity of the present solution.}. Given the Heisenberg algebra
$[\hat{x}_i,\hat{p}_j]=i\hbar\delta_{ij}\mathbb{I}$, let us first introduce the following cartesian Fock algebra,
this time in terms of the effective angular frequency $\omega$ rather than the cyclotron one, $\omega_c$,
\begin{eqnarray}
a_1 = \sqrt{\frac{m\omega}{2\hbar}}\left(\hat{x}_1+\frac{i}{m\omega}\hat{p}_1\right),&&\qquad
a^\dagger_1 = \sqrt{\frac{m\omega}{2\hbar}}\left(\hat{x}_1-\frac{i}{m\omega}\hat{p}_1\right), \nonumber \\
a_2 = \sqrt{\frac{m\omega}{2\hbar}}\left(\hat{x}_2+\frac{\gamma}{m\omega^2_0}\mathbb{I}+\frac{i}{m\omega}\hat{p}_2\right),&&\qquad
a^\dagger_2 = \sqrt{\frac{m\omega}{2\hbar}}\left(\hat{x}_2+\frac{\gamma}{m\omega^2_0}\mathbb{I}-\frac{i}{m\omega}\hat{p}_2\right),
\end{eqnarray}
which are such that,
\begin{equation}
\Big[a_i,a^\dagger_j\Big]=\delta_{ij}\,\mathbb{I}.
\end{equation}
Introduce now once again the chiral or helicity Fock algebra operators
\begin{equation}
a_\pm=\frac{1}{\sqrt{2}}\left(a_1 \mp i a_2\right),\qquad
a^\dagger_\pm=\frac{1}{\sqrt{2}}\left(a^\dagger_1 \pm i a^\dagger_2\right),
\end{equation}
\begin{equation}
\Big[a_\pm,a^\dagger_\pm\Big]=\mathbb{I}.
\end{equation}
A simple substitution then finds for the quantum Hamiltonian,
\begin{equation}
\hat{H}=\hbar\omega\left(a^\dagger_+ a_+ + a^\dagger_- a_-+1\right)
-\frac{1}{2}\hbar\omega_c\left(a^\dagger_+ a_+ - a^\dagger_- a_-\right)-\frac{\gamma^2}{2m\omega^2_0},
\end{equation}
which is thus diagonalised on the basis of Fock states $|n_-,n_+\rangle$ ($n_\pm=0,1,2,\ldots$) constructed
out of the chiral Fock algebra,
\begin{equation}
\hat{H}|n_-,n_+\rangle= E(n_-,n_+)\,|n_-,n_+\rangle,\qquad
E(n_-,n_+)=\hbar\omega_+ n_- + \hbar\omega_- n_+ + \hbar\omega-\frac{\gamma^2}{2m\omega^2_0},
\end{equation}
where,
\begin{equation}
\omega_+=\omega-\frac{1}{2}\omega_c,\qquad
\omega_-=\omega+\frac{1}{2}\omega_c.
\end{equation}

The energy eigenspectrum thus consists of normalisable and localised states. As a matter of fact, the configuration
space wave functions, $\langle x_1,x_2|n_-,n_+\rangle$, of these chiral Fock states are given as in (\ref{eq:Laguerre}),
with this time the following definition for the two variables $u$ and $\theta$,
\begin{equation}
u=\frac{m\omega}{\hbar}\left[x^2_1+\left(x_2+\frac{\gamma}{m\omega^2_0}\right)^2\right],\qquad
e^{i\theta}=\frac{x_1+i\left(x_2+\frac{\gamma}{m\omega^2_0}\right)}
{\sqrt{x^2_1+\left(x_2+\frac{\gamma}{m\omega^2_0}\right)^2}}.
\end{equation}

It is also of interest to consider the time evolution of the quantum phase space operators $(\hat{x}_i,\hat{p}_i)$
in the Heisenberg picture. Given the above expression for the quantum Hamiltonian, the time evolution of each
of the Fock algebra operators is readily identified, leading to,
\begin{eqnarray}
\hat{x}_1(t) &=& \frac{1}{2}\sqrt{\frac{\hbar}{m\omega}}\left(a_+ e^{-i\omega_- t} +a_- e^{-i\omega_+ t}
+a^\dagger_+ e^{i\omega_- t} + a^\dagger_- e^{i\omega_+ t}\right) , \nonumber \\
\hat{x}_2(t) &=& -\frac{\gamma}{m\omega^2_0}\mathbb{I}+\frac{1}{2}i\sqrt{\frac{\hbar}{m\omega}}
\left(a_+ e^{-i\omega_- t} - a_- e^{-i\omega_+ t} - a^\dagger_+ e^{i\omega_- t} + a^\dagger_- e^{i\omega_+ t}\right), \nonumber \\
\hat{p}_1(t) &=& -\frac{1}{2}im\omega\sqrt{\frac{\hbar}{m\omega}}
\left(a_+ e^{-i\omega_- t} + a_- e^{-i\omega_+ t} - a^\dagger_+ e^{i\omega_- t} - a^\dagger_- e^{i\omega_+ t}\right), \nonumber \\
\hat{p}_2(t) &=& \frac{1}{2}m\omega\sqrt{\frac{\hbar}{m\omega}}\left(a_+ e^{-i\omega_- t}
- a_- e^{-i\omega_+ t} + a^\dagger_+ e^{i\omega_- t} - a^\dagger_- e^{i\omega_+ t}\right).
\label{eq:generalsolution}
\end{eqnarray}
Of course, these expressions provide the explicit solutions to the linear Hamiltonian equations of motion
of the system, whether at the classical level, or the quantum level in the Heisenberg picture. Note well that
all the above operators $(a_\pm,a^\dagger_\pm)$ are defined by the initial Heisenberg commutation relations specified
either in the Schr\"odinger picture, or the Heisenberg picture at initial time $t=0$.

All these expressions reproduce also those of the ordinary Landau problem of Section~\ref{Sect2.1}, provided
however the limits in $\omega^2_0\rightarrow 0$ and $\gamma\rightarrow 0$ are taken appropriately.
First the linear potential term needs to be removed, $\gamma\rightarrow 0$, and only then is the
spherically symmetric well to be flattened out, $\omega^2_0\rightarrow 0$. All the expressions above
are then smoothly mapped back to those of Section~\ref{Sect2.1}. In other words, by first bringing the equilibrium
point of the total spherically symmetric potential back to the origin of the plane, $(x_1,x_2)=(0,0)$, namely by first
removing the linear contribution, and only then removing the spherical well, one reproduces the original Landau
problem.

However when the limits are considered in the reverse order, one immediately runs into singularities.
Indeed, note that when first the spherical well is flattened out while still keeping the constant force
acting on the particle, $\omega^2_0\rightarrow 0$ but $\gamma\ne 0$, singularities in the quantity $\gamma/(m\omega^2_0)$
arise and the operators $(a_2,a^\dagger_2)$, hence $(a_\pm,a^\dagger_\pm)$ are then no longer well defined. Nor is thus
the general solution in (\ref{eq:generalsolution}) and the chiral Fock states $|n_-,n_+\rangle$.

Classically, by first removing the spherical well the particle is being set free---it is not longer confined within
the well---, and being subjected to a constant force
inside the plane in conjunction with the magnetic force which is always perpendicular to its velocity, the net result is
a circular motion around a magnetic center which rather than being static as in the ordinary Landau problem, now moves at a constant
velocity in a direction perpendicular to both the magnetic field and the constant force, namely in our case along
the $x_1$ direction with the velocity,
\begin{equation}
\dot{x}^c_1=-\frac{\gamma}{B},\qquad
\dot{x}^c_2=0.
\end{equation}
In terms of the above solution considered at the classical level, in the limit $\omega^2_0\rightarrow 0$ one has $\omega_-\rightarrow 0$,
so that the actual classical solution acquires a linear time dependence---the one describing the motion of its magnetic
center at a constant velocity---combined with a periodic circular motion of angular frequency $\omega_+\rightarrow \omega_c$
once again, about that moving magnetic center. Applying a Galilei boost taking the system to the inertial frame of the magnetic center,
one recovers the ordinary Landau problem. Note that the motion of the magnetic center is along the $x_1$ axis, but with a value for
$x_2$ which is a function of the initial conditions for the classical trajectory.

Rather than considering trying to apply to the above quantum solution, in particular the construction of its chiral Fock states,
a singular limiting procedure which at the classical level produces out of the general solutions in (\ref{eq:generalsolution})
the correct ones when $\omega^2_0=0$, since the Hamiltonian equations of motion are linear and thus identical whether for
the classical or the quantum system it is more straightforward to immediately consider the situation with $\omega^2_0=0$
and $\gamma\ne 0$ at the classical level, and out of its solutions construct the appropriate realisation of the quantum
system in that singular case of the extended Landau problem. Such an approach is also simpler than trying to apply
to the ordinary Landau problem a Galilei boost from the magnetic center frame back to the initial frame in which the dynamics
of the system is being considered.

\section{The Landau Problem with a Linear Potential}
\label{Sect4}

In the symmetric gauge, the Lagrangian of the system now reads,
\begin{equation}
L=\frac{1}{2}m\left(\dot{x}^2_1+\dot{x}^2_2\right)-\frac{1}{2}B\left(\dot{x}_1 x_2 -x_1 \dot{x}_2\right)-\gamma x_2,
\end{equation}
hence the Hamiltonian,
\begin{equation}
H=\frac{1}{2m}\left(p_1+\frac{1}{2}Bx_2\right)^2+\frac{1}{2m}\left(p_2-\frac{1}{2}Bx_1\right)^2+\gamma x_2.
\label{eq:Hlinear}
\end{equation}
Trying to apply to the quantised version of the system the same types of operator redefinitions as those of
Section~\ref{Sect3} runs into the difficulty that the term linear in $\gamma x_2$ remains non diagonal in whatever
Fock state basis being considered. In order to tackle this issue, in the same way as was discussed in Section~\ref{Sect2.2}
for the ordinary Landau problem in the Landau gauge, first a canonical transformation of phase space
parametrisation is required, which readily provides at the quantum level the diagonalised Hamiltonian, hence
the solution to the quantum dynamics of the system.

Rather than specifying this canonical transformation still at the classical level and in terms of Poisson
brackets, let us already define it at the quantum level for the quantum operators and their commutation
relations in the Schr\"odinger picture, or the Heisenberg picture at time $t=0$.
Consider then the following definitions, from the variables $(\hat{x}_i,\hat{p}_i)$ to the variables
$(\hat{x}^c_1,\hat{x}^c_2,a_-,a^\dagger_-)$,
\begin{eqnarray}
\hat{x}^c_1 &=& \frac{1}{2}\hat{x}_1\,+\,\frac{1}{m\omega_c}\hat{p}_2, \nonumber \\
\hat{x}^c_2 &=& \frac{1}{2}\hat{x}_2\,-\,\frac{1}{m\omega_c}\hat{p}_1\,-\,\frac{\gamma}{m\omega^2_c}, \nonumber \\
a_- &=& \sqrt{\frac{m\omega_c}{2\hbar}}\left(\frac{1}{2}\hat{x}_1-\frac{1}{m\omega_c}\hat{p}_2\right)
\,+\,\frac{i}{m\omega_c}\sqrt{\frac{m\omega_c}{2\hbar}}\left(\hat{p}_1+\frac{1}{2}m\omega_c\hat{x}_2+\frac{\gamma}{\omega_c}\,\mathbb{I}
\right), \nonumber \\
a^\dagger_- &=& \sqrt{\frac{m\omega_c}{2\hbar}}\left(\frac{1}{2}\hat{x}_1-\frac{1}{m\omega_c}\hat{p}_2\right)
\,-\,\frac{i}{m\omega_c}\sqrt{\frac{m\omega_c}{2\hbar}}\left(\hat{p}_1+\frac{1}{2}m\omega_c\hat{x}_2+\frac{\gamma}{\omega_c}\,\mathbb{I}
\right).
\label{eq:canonicallinear}
\end{eqnarray}
The inverse relations are,
\begin{eqnarray}
\hat{x}_1 &=& \hat{x}^c_1\,+\,\sqrt{\frac{\hbar}{2m\omega_c}}\left(a_- + a^\dagger_-\right), \nonumber \\
\hat{x}_2 &=& \hat{x}^c_2\,-\,i\sqrt{\frac{\hbar}{2m\omega_c}}\left(a_- - a^\dagger_-\right), \nonumber \\
\hat{p}_1 &=& -\frac{\gamma}{\omega_c}\mathbb{I}\,-\,\frac{1}{2}m\omega_c\hat{x}^c_2\,-\,\frac{1}{2}im\omega_c\sqrt{\frac{\hbar}{2m\omega_c}}
\left(a_- - a^\dagger_-\right), \nonumber \\
\hat{p}_2 &=& \frac{1}{2}m\omega_c \hat{x}^c_1\,-\,\frac{1}{2}m\omega_c\sqrt{\frac{\hbar}{2m\omega_c}}
\left(a_- + a^\dagger_-\right).
\end{eqnarray}
It then follows that the two sectors $(\hat{x}^c_1,\hat{x}^c_2)$ and $(a_-,a^\dagger_-)$ commute with one another, while
we have,
\begin{equation}
\Big[\hat{x}^c_1,\hat{x}^c_2\Big]=-\frac{i\hbar}{B}\,\mathbb{I},\qquad
\Big[a_-,a^\dagger_-\Big]=\mathbb{I}.
\end{equation}
Hence indeed this reparametrisation of phase space is a canonical transformation preserving canonical Poisson brackets
(one may rescale, say $\hat{x}^c_2$, by the factor $(-B=-m\omega_c)$, if ones prefers).
Related to the magnetic center sector, $(\hat{x}^c_1,\hat{x}^c_2)$,
we also have the following Fock algebra generators,
\begin{equation}
\hat{x}^c_1=\sqrt{\frac{\hbar}{2m\omega_c}}\left(a_+ + a^\dagger_+\right),\qquad
\hat{x}^c_2=i\sqrt{\frac{\hbar}{2m\omega_c}}\left(a_+ - a^\dagger_+\right),
\end{equation}
which are such that,
\begin{equation}
\Big[ a_+, a^\dagger_+\Big]=\mathbb{I}.
\end{equation}
These operators $(a_+,a^\dagger_+)$ correspond to the right-handed chiral mode of the ordinary Landau problem
which in that context has no time dependence, but acquires one in the present case because of the constant force
of strength $\gamma$ which indeed sets into motion the magnetic center. Note that $\hat{x}^c_1$ corresponds to
the magnetic center position along the $x_1$ axis, while the contribution $\hat{x}^c_2$
to $\hat{x}_2$ corresponds to its position along the $x_2$ axis. The contribution $(-\gamma/\omega_c=-m\gamma/B)$
to $\hat{p}_1$ corresponds to the velocity momentum of the magnetic center, $m\dot{x}^c_1$. Finally, $(a_-,a^\dagger_-)$
correspond to the left-handed chiral rotating mode with angular frequency $\omega_c$ of the ordinary Landau problem
in the magnetic center inertial frame

A direct substitution of these relations in the Hamiltonian finds,
\begin{equation}
\hat{H}=\hbar\omega_c\left(a^\dagger_- a_- +\frac{1}{2}\right)\,+\,\gamma\hat{x}^c_2\,+\,
\frac{1}{2}m\left(\frac{\gamma}{B}\right)^2.
\label{eq:Hlinear2}
\end{equation}
The physical meaning of each of these contributions should be clear enough. The very last term corresponds
to the kinetic energy of the magnetic center moving at constant velocity of norm $|\gamma|/B$. The term before the
last represents the potential energy along the $x_2$ direction, $\gamma x_2$, of the magnetic center
position along that axis. And finally the very first
contribution with the two terms in parentheses measures the excitation energy of the usual Landau levels of the
ordinary Landau problem, as seen from the magnetic center inertial frame.

This expression for the quantum Hamiltonian also makes it clear which basis of states diagonalises that
operator, hence solves the quantum dynamics of the system. Given the Fock states $|n_-\rangle$ associated to the
$(a_-,a^\dagger_-)$ Fock algebra, and position eigenstates, $\hat{x}^c_2|x^c_2\rangle=x^c_2|x^c_2\rangle$, for the magnetic center position
along the $x_2$ axis, the basis of the space of quantum states which diagonalises the dynamics is spanned
by the states $|n_-,x^c_2\rangle$ ($n_-=0,1,2,\ldots$, $x^c_2\in\mathbb{R}$), with the normalisation,
\begin{equation}
\langle n_-,x^c_2|m_-,{x'}^c_2\rangle=\delta_{n_-,m_-}\,\delta(x^c_2 - {x'}^c_2).
\end{equation}
One has,
\begin{equation}
\hat{H}|n,x^c_2\rangle=E(n,x^c_2)|n,x^c_2\rangle,\qquad
E(n,x^c_2)=\hbar\omega_c\left(n+\frac{1}{2}\right)+\gamma x^c_2+\frac{1}{2}m\left(\frac{\gamma}{B}\right)^2.
\label{eq:linearspectrum}
\end{equation}

The fact that the magnetic center component of these quantum states is not normalisable makes now perfect
physical sense. Indeed, the motion of that magnetic center is that of a free particle with a predetermined
velocity set by the ratio $(-\gamma/B)$ in the $x_1$ direction. Hence in the configuration space representation
states possess wave functions with a plane wave component in that direction, which is not normalisable,
and leads to the above $\delta$ function normalisation for energy eigenstates. Having chosen from the outset not
to work in the Landau gauge guarantees without ambiguity that this lack of normalisability is indeed related
to a physical feature of the solution rather than a not totally appropriate choice of phase space parametrisation.

Given the Hamiltonian, it is also possible to determine the time dependence of the phase space operators in the
Heisenberg picture. One finds,
\begin{eqnarray}
\hat{x}_1(t) &=& \hat{x}^c_1\,-\,\frac{\gamma}{B}t\,\mathbb{I}\,+\,\sqrt{\frac{\hbar}{2m\omega_c}}
\left(a_- e^{-i\omega_c t} + a^\dagger_- e^{i\omega_c t}\right), \nonumber \\
\hat{x}_2(t) &=& \hat{x}^c_2\,-\,i\sqrt{\frac{\hbar}{2m\omega_c}}
\left(a_- e^{-i\omega_c t} - a^\dagger_- e^{i\omega_c t}\right), \nonumber \\
\hat{p}_1(t) &=& -\frac{\gamma}{\omega_c}\,\mathbb{I}\,-\,\frac{1}{2}m\omega_c\,\hat{x}^c_2\,-\,\frac{1}{2}im\omega_c
\sqrt{\frac{\hbar}{m\omega_c}}\left(a_- e^{-i\omega_c t} - a^\dagger_- e^{i\omega_c t}\right), \nonumber \\
\hat{p}_2(t) &=& \frac{1}{2}m\omega_c\,\hat{x}^c_1\,-\,\frac{1}{2}\gamma t\,\mathbb{I}\,-\,
\frac{1}{2}m\omega_c\sqrt{\frac{\hbar}{m\omega_c}}\left(a_- e^{-i\omega_c t} - + a^\dagger_- e^{i\omega_c t}\right),
\end{eqnarray}
namely,
\begin{equation}
\hat{x}^c_1(t)=\hat{x}^c_1-\frac{\gamma}{B}t\,\mathbb{I},\qquad
\hat{x}^c_2(t)=\hat{x}^c_2.
\end{equation}
In the expressions for $\hat{x}_i(t)$ one may recognise the solution to the ordinary Landau problem (the terms involving
$a_-$ and $a^\dagger_-$), valid in the magnetic center inertial frame, to which is added the Galilei boost
with the constant velocity of the magnetic center towards the inertial frame with the potential energy $\gamma x_2$, and the
initial position of that magnetic center along both the $x_1$ and $x_2$ axes. Incidentally, and as was indicated
at the end of the previous Section, this is in fact how the change of variables (\ref{eq:canonicallinear}) was identified initially.
Writing out the classical solution for $x_i(t)$ precisely in that way, and then identifying what are the ensuing
expressions for $p_i(t)$ given that
\begin{equation}
p_1(t)=m\dot{x}_1(t)+\frac{1}{2}Bx_2(t),\qquad
p_2(t)=m\dot{x}_2(t)-\frac{1}{2}Bx_1(t),
\end{equation}
all in a manner that meets all Hamiltonian equations of motion, the appropriate operators in the Heisenberg picture
are identified. Upon substitution into the quantum Hamiltonian, one then is bound to find the quantum solution for it as well.

Note also that in the same spirit as that of the entire discussion so far, the above solution of the singular
Landau problem extended with a linear potential has remained purely algebraic, without the need to solve
the differential Schr\"odinger wave equation. To the best of our knowledge, this specific approach and
construction for the extended singular Landau problem is not available in the literature.

As a passing remark of some interest as well, note that given a projection onto any subspace of Hilbert space
corresponding to a Landau sector at fixed level $n_-$, namely onto the subspace spanned by the states $|n_-,x^c_2\rangle$
for a fixed $n_-$ and for all $x^c_2\in\mathbb{R}$, in effect the only remaining degrees of freedom are those
of the magnetic center, $\hat{x}^c_1$ and $\hat{x}^c_2$, which obey the commutation relation of the ordinary Moyal--Voros
plane of noncommutative geometry\cite{scholtz,scholtz2}, $[\hat{x}^c_1,\hat{x}^c_2]=-i(\hbar/B)\mathbb{I}$.
This result is readily established through the present discussion without the need of any actual calculation
of projected matrix elements, in contradistinction to the usual derivation of this result available in the literature\cite{macris}.

Finally, it now becomes feasible without much difficulty to explicitly work out the configuration
space wave functions for all energy eigenstates, given the expressions of the operators $(\hat{x}^c_1,\hat{x}^c_2,a_-,a^\dagger_-)$
in terms of $(\hat{x}_i,\hat{p}_i)$. This task would rather be a great deal more involved were one to consider from the outset the
differential Schr\"odinger equation eigenvalue problem given the Hamiltonian in the form of (\ref{eq:Hlinear}).
Without going here into the details of the calculation, let us only mention that in a first step one works out the
wave function for the lowest Landau sector, $|n_-=0,x^c_2\rangle$, as a function of $x^c_2$. Applying then the
operator $a^\dagger_-$ onto that solution, one readily finds the wave functions for all states $|n_-,x^c_2\rangle$, including
their normalisation. When normalising the position eigenstates of the configuration space basis as is usual,
\begin{equation}
\langle x_1,x_2|x'_1,x'_2\rangle=\delta(x_1-x'_1)\,\delta(x_2-x'_2),\qquad
\hat{x}_i|x_1,x_2\rangle=x_i\,|x_1,x_2\rangle,
\end{equation}
one finds,
\begin{eqnarray}
\langle x_1,x_2|n_-,x^c_2\rangle &=& \left(\frac{m\omega_c}{\pi\hbar}\right)^{1/4}\,
\left(\frac{m\omega_c}{2\pi\hbar}\right)^{1/2}\,\frac{(-i)^n}{\sqrt{2^n\,n!}}\,\times \nonumber \\
&&\times\,e^{-i\frac{m\omega_c}{\hbar} x_1(x^c_2-\frac{1}{2}x_2+\frac{\gamma}{m\omega^2_c})}\,
e^{-\frac{m\omega_c}{2\hbar}(x_2-x^c_2)^2}\,H_n\left(\sqrt{\frac{m\omega_c}{\hbar}}(x_2-x^c_2)\right),
\label{eq:wave1}
\end{eqnarray}
$H_n(u)$ being the Hermite polynomial of order $n$.

Hence these energy eigenstates are localised only in the $x_2$ direction, while in the $x_1$ direction they
are totally delocalised and non normalisable, since they propagate in time in that direction as a free particle
of predetermined constant velocity $(-\gamma/B)$. The probability density of these states thus also looks like
a series of $(n+1)$ parallel stripes with exponentially smooth edges, and invariant under translations along the $x_1$ axis.

Having constructed a complete solution of this singular Landau problem extended by a linear potential,
note how all these results are smooth in the parameter $\gamma$, and indeed reproduce in the limit $\gamma\rightarrow 0$
those of the ordinary Landau problem. Clearly in that limit the states $|n_-,x^c_2\rangle$ remain
non normalisable, because the right-handed chiral sector $(a_+,a^\dagger_+)$ has been diagonalised rather
in terms of the operator $\hat{x}^c_2$, namely by having required the magnetic center position to be sharp
in the $x_2$ direction, hence totally delocalising its position along $x_1$, since the two coordinates of the
magnetic center obey a Heisenberg algebra and do not commute. However in the limit $\gamma=0$ each of the Landau sectors
distinguished by $n_-$ becomes once again energy degenerate, allowing for another choice of energy eigenstate basis in the magnetic
center sector. Choosing for it once again the orthonormalised Fock state basis $|n_-,n_+\rangle$, one recovers precisely
exactly the same solution as that constructed in Section~\ref{Sect2.1} (with $(\overline{x}_1,\overline{x}_2)=(0,0)$).
However the construction of the present Section is useful even for the ordinary Landau problem, when a sharp rectilinear
edge is introduced in the plane, as we now discuss.

\section{The Landau Problem in the Half-Plane with a Linear Potential}
\label{Sect5}

A noteworthy feature of the energy spectrum (\ref{eq:linearspectrum}) is that it is unbounded below, and
yet the quantum system (as well as the classical system) remains stable because the magnetic force combines with
the constant force of strength $\gamma$ to keep the particle rotating periodically around a magnetic center
that moves at constant velocity along the $x_1$ axis. Clearly, the reason for this unboundedness in the energy
is that the particle may ``fall off the plane" in one of the $x_2$ directions, so to say, a way of putting this fact which
is particularly appropriate in case the constant force is indeed that of gravity.

A manner in which to avoid this unboundedness is to restrict the range of $x_2$, namely consider now the Landau problem
on the half-plane with still the linear potential. Assuming now that $\gamma>0$, let us therefore restrict to the $x_2\ge X_2$
half-plane for some value of $X_2\in\mathbb{R}$, and reconsider the solution of the quantised system.
Such a situation is also of physical interest.
Given that the Landau problem is of relevance to the quantum Hall effect, in particular in its integer and even
fractional manifestations, combining the magnetic field with a constant force acting inside the plane, be it
electrical or gravitational, may allow for interesting properties of that collective quantum fermion phenomenon
to manifest themselves. In the gravitational context by tilting the quantum Hall device towards the vertical direction,
one is setting up a gravitational quantum well in combination with the quantum Hall effect. Given that the energy
quantisation of gravitational quantum states in a gravitational well has been observed already with ultra-cold
neutrons\cite{Nesv}, a quantum Hall set-up may provide an interesting alternative to such measurements, provided
the orders of magnitude for any effect are large enough to be observable. Of course, given the weakness of gravity,
an electric field stands a much better chance to display any such interesting effects.

Clearly the change of variable specified in (\ref{eq:canonicallinear}) is still in order in this case,
leading to the Hamiltonian in (\ref{eq:Hlinear2}). However there is a subtlety now, given the boundary at
$x_2=X_2$. Since one has to restrict now to the quantum Hilbert space of configuration space wave functions
that vanish at that boundary as well as inside the excluded domain, $x_2<X_2$, the conjugate momentum operator
$\hat{p}_2$ does not possess a self-adjoint extension, and is in fact only symmetric on that space\cite{Bonneau}. Indeed,
as the differential operator $(-i\hbar\partial_2)$, the operator $\hat{p}_2$ maps quantum states out of that
Hilbert space, while we have, for any two states $|\psi\rangle$ and $|\varphi\rangle$ represented by functions
$\psi(x_2)$ and $\varphi(x_2)$ in the $(\hat{x}_2,\hat{p}_2)$ sector,
\begin{eqnarray}
\langle\varphi|\hat{p}_2\psi\rangle &=& \int_{X_2}^{+\infty} dx_2\,\varphi^*(x_2)\left(-i\hbar\frac{d}{dx_2}\psi(x_2)\right) \nonumber \\
&=& -i\hbar\,\int_{X_2}^{+\infty}\,d\left(\varphi^*(x_2)\psi(x_2)\right)\,+\,
\int_{X_2}^{+\infty}dx_2\left(-i\hbar\frac{d}{dx_2}\varphi(x_2)\right)^*\,\psi(x_2) \nonumber \\
&=& -i\hbar\,\int_{X_2}^{+\infty}\,d\left(\varphi^*(x_2)\psi(x_2)\right)\,+\,\langle\hat{p}_2\varphi|\psi\rangle.
\end{eqnarray}
Given that both wave functions $\psi(x_2)$ and $\varphi(x_2)$ mush vanish at $x_2=X_2$ and $x_2\rightarrow +\infty$
(the latter condition applies since states must be normalisable at least in the $x_2$ direction),
it follows that $\hat{p}_2$ is indeed a symmetric operator.

However among those operators contributing to the quantum Hamiltonian still given as in (\ref{eq:Hlinear2}),
this ambiguity affects only the $(a_-,a^\dagger_-)$ operators, which are thus no longer
adjoints of one another, since each maps outside the considered space of quantum states. However, the
other operator involved in diagonalising the Hamiltonian, namely $\hat{x}^c_2$, is not affected by
that lack of self-adjointness in $\hat{p}_2$ since it is given as
\begin{equation}
\hat{x}^c_2=\frac{1}{2}\hat{x}_2-\frac{1}{m\omega_c}\hat{p}_1-\frac{\gamma}{m\omega^2_c},
\end{equation}
which is an expression that does not involve the operator $\hat{p}_2$, in contradistinction to the operators $(a_-,a^\dagger_-)$.
Consequently, one may still consider the space of eigenstates of $\hat{x}^c_2$, whose wave functions are
given as in the construction of the previous Section, as a factor in the tensor product structure providing
a basis of energy eigenstates. For the remaining separable factor in that tensor product, even though one may no longer
exploit the existence of a Fock vacuum annihilated by $a_-$ in order to diagonalise the Hamiltonian,
it is rather the latter Hamiltonian which needs diagonalisation, a procedure which does not necessarily require
a construction of Fock states representing a Fock algebra which in the present case does not exist. As will be seen
hereafter, in contradistinction to the $(a_-,a^\dagger_-)$ operators, the Hamiltonian operator
itself is not affected by that issue and does possess a self-adjoint extension\cite{Bonneau}.

Writing the quantum Hamiltonian in the form,
\begin{equation}
\hat{H}=\frac{1}{2}\hbar\omega_c\left(a^\dagger_- a_- + a_- a^\dagger_-\right) + \gamma\hat{x}^c_2
+\frac{1}{2}m\left(\frac{\gamma}{B}\right)^2,
\end{equation}
and using the explicit expressions for $a_-$ and $a^\dagger_-$ in (\ref{eq:canonicallinear}), one undoes
part of the canonical transformation to find,
\begin{equation}
\hat{H}=\frac{1}{2}m\omega^2_c\left(\frac{1}{2}\hat{x}_1-\frac{1}{m\omega_c}\hat{p}_2\right)^2\,+\,
\frac{1}{2m}\left(\hat{p}_1+\frac{1}{2}m\omega_c\hat{x}_2+\frac{\gamma}{\omega_c}\right)^2\,+\,
\gamma\hat{x}^c_2\,+\,\frac{1}{2}m\left(\frac{\gamma}{B}\right)^2.
\end{equation}
Let us now consider the diagonalisation of this operator by working in the configuration space
wave function representation for quantum states, $\psi(x_1,x_2)$. Since energy eigenstates are
certainly eigenstates of $\hat{x}^c_2$, their wave functions certainly separate as
\begin{equation}
\psi_{E,x^c_2}(x_1,x_2)=e^{-i\frac{m\omega_c}{\hbar}x_1(x^c_2-\frac{1}{2}x_2+\frac{\gamma}{m\omega^2_c})}\,
\varphi_{E,x^c_2}(x_2),
\end{equation}
$E$ denoting their energy eigenvalue. A direct substitution in the stationary Schr\"odinger equation
in the configuration space representation then reduces to, for any given value of $x^c_2$,
\begin{equation}
\left\{-\frac{\hbar}{2m}\frac{d^2}{d x^2_2}\,+\,\frac{1}{2}m\omega^2_c\left(x_2-x^c_2\right)^2\,+\,\gamma x^c_2\,+
\frac{1}{2}m\left(\frac{\gamma}{B}\right)^2\right\}\,\varphi_{E,x^c_2}(x_2)= E\,\varphi_{E,x^c_2}(x_2),
\label{eq:Schro}
\end{equation}
where one must also meet the condition $\varphi_{E,x^c_2}(x_2=X_2)=0$ which will imply a quantisation rule
for the energy values $E$. Note that indeed this operator possesses a self-adjoint extension for this
choice of boundary conditions\cite{Bonneau}.

Introducing the notations,
\begin{equation}
u=\sqrt{\frac{2m\omega_c}{\hbar}}\left(x_2 - x^c_2\right),\qquad
a=-\frac{1}{\hbar\omega_c}\left(E-\gamma\,x^c_2\,-\,\frac{1}{2}m\left(\frac{\gamma}{B}\right)^2\right),
\end{equation}
the above eigenvalue equation becomes
\begin{equation}
\left(\frac{d^2}{du^2}\,-\,\left(\frac{1}{4}u^2+a\right)\right)\varphi_{E,x^c_2}(u)=0.
\label{eq:UV}
\end{equation}
The general solution to this equation is a linear combination of the two parabolic cylinder functions
$U(a,u)$ and $V(a,u)$\cite{AS}. However since wave functions are required to vanish at $x_2\rightarrow+\infty$,
only the $U(a,u)$ branch is allowed\footnote{Both functions $U(a,u)$ and $V(a,u)$ diverge as $u\rightarrow -\infty$,
unless $a=-n-1/2$ for $n=0,1,2,\ldots$ in which case only $U(a,u)$ also vanishes in that limit, and in fact
reduces\cite{AS} to $U(-n-1/2,u)=2^{-n/2}e^{-u^2/4}H_n(u/\sqrt{2})$.}. Hence the solution is of the form,
\begin{equation}
\varphi_{E,x^c_2}(x_2)=N(E,x^c_2)\,U(a,u),
\end{equation}
$N(E,x^c_2)$ being some normalisation factor. Consequently the energy quantisation condition is given
by the boundary condition,
\begin{equation}
U\left(a,\sqrt{\frac{2m\omega_c}{\hbar}}(X_2-x^c_2)\right)=0.
\label{eq:E}
\end{equation}
Even though an explicit resolution of this condition requires a numerical analysis, given (\ref{eq:UV})
it should be clear that this condition implies that the spectrum of $a$ values belongs to a semi-infinite discrete set labelled as
\begin{equation}
a=-a_n\left(X_2-x^c_2\right),\qquad n=0,1,2,\ldots,
\end{equation}
where each of the quantities $a_n(X_2-x^c_2)$ is a continuous function of $(X_2-x^c_2)$, while altogether they define a set of
increasing values as $n$ increases. Indeed, when multiplied
by a factor $(-1)$ and up to normalisation factors, (\ref{eq:UV}) is the Schr\"odinger wave equation for a
harmonic oscillator with eigenvalue $(-a)$, of which the quadratic potential, namely $u^2/4$,
is truncated away for $u<\sqrt{2m\omega_c/\hbar}(X_2-x^c_2)$.
Since this potential with an infinite wall at $u=\sqrt{2m\omega_c/\hbar}(X_2-x^c_2)$
is bounded below and unbounded above, its spectrum of standing waves and energy eigenvalues is certainly both
bounded below and semi-infinite discrete, with growing eigenvalues $(-a)$ as the level index
quantum number $n=0,1,2,\ldots$ keeps increasing.

Note however that the energy quantisation condition for $a$, hence its spectrum of values $a_n(X_2-x^c_2)$ is independent of
the parameter $\gamma$. For instance, since the function $U(a,u)$ vanishes in the limit $u\rightarrow -\infty$
only provided\cite{AS} $a=-n-1/2$, one has,
\begin{equation}
\lim_{X_2\rightarrow -\infty}a_n\left(X_2-x^c_2\right)=n+\frac{1}{2},\qquad
\lim_{x^c_2\rightarrow +\infty}a_n\left(X_2-x^c_2\right)=n+\frac{1}{2}.
\end{equation}

In conclusion, the configuration space wave functions of the energy eigenstates of the system are given
in the form,
\begin{equation}
\psi_{n,x^c_2}(x_1,x_2)=N(n,x^c_2)\,e^{-i\frac{m\omega_c}{\hbar}x_1(x^c_2-\frac{1}{2}x_2+\frac{\gamma}{m\omega^2_c})}\,
U\left(-a_n\left(X_2-x^c_2\right),\sqrt{\frac{2m\omega_c}{\hbar}}\left(x_2-x^c_2\right)\right),
\label{eq:wave2}
\end{equation}
$N(n,x^c_2)$ being a normalisation constant to be determined, while the energy spectrum is,
\begin{equation}
E(n,x^c_2)=\hbar\omega_c\,a_n\left(X_2-x^c_2\right)\,+\,\gamma\,x^c_2\,+\,\frac{1}{2}m\left(\frac{\gamma}{B}\right)^2.
\label{eq:spectrum2}
\end{equation}

As compared to the results obtained in the plane in Section~\ref{Sect4}, the only difference is the replacement
by the quantities $a_n(X_2-x^c_2)$ of the contributions in $(n+1/2)$ of the left-handed chiral mode $(a_-,a^\dagger_-)$,
while in the wave functions of these energy eigenstates the Hermite polynomial contribution multiplied by the Gaussian factor is replaced
by that of the parabolic cylinder function. Furthermore there is no restriction whatsoever on the possible values
for $x^c_2$, even though the particle remains confined to the $x_2\ge X_2$ region. Yet the energy spectrum remains now
bounded below.

Note that in the limit where the edge of the half-plane is pushed out again back to infinity, namely $X_2\rightarrow -\infty$,
the above results reduce smoothly back to those of Section~\ref{Sect4}, as they should. However for any finite position
of the edge at $x_2=X_2$, even in the limit when $\gamma\rightarrow 0^+$, the energy spectrum remains non degenerate
since the values $a_n(X_2-x^c_2)$ are functions of $(X_2-x^c_2)$, hence of $x^c_2$, and are independent of $\gamma$.
In other words, the Landau level degeneracies of the ordinary Landau problem in the plane are lifted because
of the interactions brought about by the edge---namely an infinite potential wall---at a finite position in the plane.
Incidentally, the set of transformations in (\ref{eq:canonicallinear}), identified first by considering the
extra linear potential added to the ordinary Landau problem, proved essential in being able to construct the above
explicit solution for the ordinary Landau problem in the half-plane.

Some features of the above complete solution may also be understood from the point of view of the classical
trajectories in the half-plane, even in the presence or not of the constant force of strength $\gamma$. Given any value for
$x^c_2>X_2$ however close to $X_2$, there always exist solutions of sufficiently small energy such that the
radius of their circular motion about their (possibly moving, if $\gamma\ne 0$) magnetic center
remains less than the difference $(x^c_2-X_2>0)$, so that the particle then does not bounce off the wall at $x_2=X_2$.
Such trajectories are not distorted by the presence of the edge. At the quantum level because of the nonlocal nature
of their wave function, such states display a slight deformation of their wave function, hence also of their energy value,
but the less so the less is their energy and the larger is the value for $x^c_2$ away from $X_2$.
However as soon as the energy of the classical trajectory becomes large enough so that its radius becomes larger than
$(x^c_2-X_2)$, the particle starts bouncing periodically off the wall at $x_2=X_2$ in a series of elastic collisions,
and is, in effect, set into motion---in case $\gamma\ne 0$ this extra motion is superposed to that of the magnetic center already---along the
$x_1$ axis in the positive direction. These trajectories are thus distorted, and even more so are the quantum
states associated to such values of $x^c_2$ and energy. Finally even when $x^c_2$ lies inside the ``forbidden" region,
$x^c_2<X_2$, there do exist classical solutions of sufficiently large radius, namely energy, hence also quantum states
of sufficiently large energy. But these states suffer the strongest distortion in wave function and energy values
away from the equally spaced energy spectrum when the infinite wall at $x_2=X_2$ is absent.

In terms of the effective harmonic potential contributing in the Schr\"odinger equation
for $\varphi_{E,x^c_2}(x_2)$ in (\ref{eq:Schro}), namely
\begin{equation}
V_{\rm eff}=\frac{1}{2}m\omega^c_2\left(x_2-x^c_2\right)^2\ \ {\rm for}\ \ x_2\ge X_2;\qquad
V_{\rm eff}=+\infty\ \ {\rm for}\ \ x_2<X_2,
\end{equation}
which is thus truncated away on the left-hand side for $x_2< X_2$, the three typical situations discussed above
correspond to when the minimum of that potential at $x_2=x^c_2$ lies, respectively, well inside the
region $x_2>X_2$, or close to the edge, and finally inside the forbidden region $x_2<X_2$. Solutions then
correspond to standing waves inside this truncated harmonic well, which need to vanish at the infinite wall.
As was note previously, this is also the reason why the spectrum of $a_n(X_2-x^c_2)$ values is always bounded below
and discrete, as confirmed by a numerical analysis, and depends on $x^c_2$ in such a manner that the total
energy spectrum (\ref{eq:spectrum2}) remains bounded below however large and negative $x^c_2$ may be.

\section{Conclusions}
\label{SectConclusions}

This paper considered different variations on the same theme of the ordinary Landau problem. By first
understanding why the choice of Landau gauge for the magnetic vector potential leads to non countable and
non normalisable energy eigenstates whereas a solution in any other gauge produces an energy eigenbasis of 
countable and normalisable eigenstates, different canonical transformations of the phase space variables
have been discussed enabling a straightforward resolution of different extensions of the Landau problem
with a potential energy quadratic and linear in the plane cartesian coordinates.

As a matter of fact the main aim of this study was the explicit resolution solely through algebraic means
of a singular extension of the Landau problem, namely when a potential energy only linear in the cartesian
coordinates is introduced. Based on a specific canonical transformation, a clear separation of magnetic
center degrees of freedom and chiral rotating ones is achieved, allowing for a simple identification of
the energy eigenspectrum and even of all its configuration space wave functions. The advantages of this approach
are then finally brought to bear on the explicit and analytic solution of this singular Landau problem in the half-plane.

In order to come closer to an actual physical situation of physical interest in the quantum Hall context,
whether the linear potential is related to a constant electric field or a gravitational well, the present study
may be pursued by adding three more edges in the half-plane in order to build up a slab of finite
extent, as a model for an actual experimental device for quantum Hall measurements. As a consequence,
presumably the choice of variables which enabled the present solutions in the plane and in the half-plane
will no longer be the most appropriate. Indeed, looking at the wave functions in (\ref{eq:wave1}) and (\ref{eq:wave2}),
in that form it does not appear possible to enforce a vanishing wave function at two separate values of $x_1$.
This is due to the fact that the plane wave component in $x_1$ of these wave functions derives from the eigenvalue
value equation for the magnetic center coordinate $\hat{x}^c_2$, which also contributes linearly to the
Hamiltonian. However, since the two magnetic center coordinates $\hat{x}^c_1$ and $\hat{x}^c_2$ do not commute,
one cannot restrict both to be sharp, and by restricting the value of wave functions for specific values of $x_1$
certainly implies that energy eigenstates are no longer eigenstates of $\hat{x}^c_2$. In other words,
like in the half-plane where one could no longer first diagonalise the Fock algebra and then produce
the energy eigenspectrum, in a finite slab neither the Fock algebra nor the operator $\hat{x}^c_2$ may be
diagonalised together with the Hamiltonian. From that point of view the diagonalisation of the Hamiltonian
in the form of (\ref{eq:Hlinear}), or possibly even in the Landau gauge rather than the symmetric gauge,
certainly looks more appropriate. This issue deserves a dedicated study, which is likely to lead once again to
parabolic cylinder functions in the $x_2$ direction, and ordinary trigonometric standing waves in the $x_1$
direction.

However, for what concerns the energy spectrum one should not expect a result that differs much qualitatively from
that in (\ref{eq:spectrum2}). Namely besides a term analogous to the one linear in $\gamma x^c_2$,
there should be another contribution whose scale is set by the Landau problem itself, $\hbar\omega_c$. Furthermore,
this latter contribution should remain independent of the coefficient setting the strength of the linear potential, $\gamma$.
Consequently, the only effect of introducing this extra interaction energy in the system is to slightly tilt the spectrum
of Landau levels.

To assess under which experimental conditions such effects may become observable, one needs to understand how
much the energy spectrum---namely that of the density of states in conduction properties
of such Hall probes---is function of the geometry of such a finite slab as compared to the effects of a linear
potential term, as may be induced by an electric field or the gravitational interaction. But whatever the finer
details of these dependencies turn out to be, it remains a fact that the factors setting the scales for these two types
of effects are the Landau level gap, $\hbar\omega_c$, and the potential energy $\gamma x_2$. This should allow
for first order estimates already.

Beyond the analysis considered here, another extension of potential interest is to include the spin 1/2
degrees of freedom of the charged particle, as is indeed the case for electrons in actual quantum Hall experiments.

Finally, motivated by totally different considerations, it would also be interesting to reconsider the
present singular Landau problem in the plane and the half-plane in the context of noncommutative quantum
mechanics, namely for the Landau problem defined over the Moyal--Voros plane, using the techniques
developed in Refs.\cite{scholtz2,Moyal,Ben} and having in mind to possibly identify some approach
to experimentally set upper bounds on the noncommutativity parameter of noncommutative space(time) geometry.

\section*{Acknowledgements}

This work was initiated while two of us (JG and MNH) were visiting the Physics Department of the University at
Zambia in April 2008. It was completed during the International Workshop on Coherent States, Path Integrals and
Noncommutative Geometry, held at the National Institute for Theoretical Physics (NITheP, Stellenbosch, South
Africa) on 4--22 May 2009. JG and MNH wish to thank Prof. H. V. Mweene and the Physics Department at the
University of Zambia for a most pleasant and warm welcome, and for financial support towards both their trip
and their stay. All three authors are also grateful to the organisers of the above Workshop as well as NITheP
and its Director, Prof. Frederik Scholtz,  for the financial support having made their participation possible,
and for NITheP's always warm, inspiring and wonderful hospitality. They wish to thank Dr. Joseph Ben Geloun, and
Profs. Frederik Scholtz and Bo-Sture Skagerstam for their interest in this work and for useful discussions during the Workshop.

JG acknowledges the Abdus Salam International Centre for Theoretical Physics (ICTP, Trieste, Italy)
Visiting Scholar Programme in support of a Visiting Professorship at the ICMPA-UNESCO (Republic of Benin).
The work of JG is supported in part by the Institut Interuniversitaire des Sciences Nucl\'eaires (I.I.S.N., Belgium),
and by the Belgian Federal Office for Scientific, Technical and Cultural Affairs through
the Interuniversity Attraction Poles (IAP) P6/11.

\end{document}